\documentstyle[aps,prb,preprint,amsmath,tighten]{revtex}
\begin{document}
\draft \title{Identifying Proteins of High Designability via
Surface-Exposure Patterns}
\author{Eldon G. Emberly$^{1}$, Jonathan Miller$^1$, Chen Zeng$^2$,
 Ned S. Wingreen$^1$, and Chao Tang$^{1,*}$}
\address{$^1$NEC Research Institute, 4 Independence Way, Princeton, NJ
08540, USA \linebreak
$^2$ Department of Physics, George Washington University,
Washington, D.~C.~20052, USA \linebreak
$^*$ Corresponding author: email: tang@research.nj.nec.com P:
(609) 951-2644 F: (609) 951-2496}
\maketitle
\bigskip
\pacs{Keywords: protein design, protein structure prediction, 
off-lattice model, hydrophobicity}
\pagebreak
\section*{abstract}
Using an off-lattice model, we fully enumerate folded
conformations of polypeptide chains of up to $N = 19$ monomers.
Structures are found to differ markedly in {\it designability},
defined as the number of sequences with that structure as a unique
lowest-energy conformation.  We find that designability is closely
correlated with the pattern of surface exposure of the folded
structure.  For longer chains, complete enumeration of structures is
impractical.  Instead, structures can be randomly sampled, and
relative designability estimated either from designability within
the random sample, or directly from surface-exposure pattern.  We
compare the surface-exposure patterns of those structures identified
as highly designable to the patterns of naturally occurring proteins.
\pagebreak
\section{Introduction}
Naturally occurring proteins fold into specific three-dimensional
structures to achieve their unique functionality.\cite{Creighton93}
For many proteins, it has been shown that the amino-acid sequence
alone is sufficient to determine the folded
conformation.\cite{Anfissen} Interestingly, out of all geometrically
possible folds, nature seems to have selected only a small set of fold
families.\cite{Chothia92,Orengo94,Brenner97,Govin99} This selection
may arise, in part, from differences in the {\it designability} of
folded structures.\cite{Fink87,Yue95,Govin95,Li96} By definition, the
designability of a structure is the number of amino-acid sequences
with that structure as the lowest-free-energy conformation.  In
lattice models, where it is possible to enumerate all compact
structures, there exists a small class of highly designable
structures, {\it i.e.}  structures which are unique lowest-energy
conformations of many more than their share of
sequences.\cite{Li96,Li98} The sequences associated with these highly
designable structures are found to have protein-like properties:
mutational stability,\cite{Govin95,Li96} thermodynamic
stability,\cite{Li96,Li98} and fast folding kinetics.\cite{Melin99}
The topology of the neutral networks formed by the sequences of
designable lattice model structures have also received
study.\cite{Bauer} Recently, off-lattice studies of protein structures
have also shown that certain backbone configurations are highly
designable, and that the associated sequences have enhanced mutational
and thermodynamic stability.\cite{Miller00} Therefore, whether one's
goal is to better understand existing protein fold families or to
design novel folds,\cite{Miller00} designability may offer a way to
identify structures and sequences with protein-like folding
properties.

In previous work, the determination of a structure's designability has
relied upon the enumeration of a wide cross section of all possible
structures.  This is because the designability of one structure
depends on competition for sequences from other structures.  For short
chains on lattices it is straightforward to enumerate all compact
structures.\cite{Lau89,Chan90,Shak90,Camacho93,Pande94,Govin95,Li96}
For off-lattice models, one
approach has been to enumerate all structures obtainable with a small,
discrete set of backbone dihedral angles.\cite{Miller00}
Clearly, for long peptide chains, this complete enumeration is
infeasible, even for a small set of dihedral angles.
Can one nevertheless identify highly designable long-chain
protein structures?

In this paper, we present evidence from studies of short chains, up to
$N=19$, that the designability of a structure can be predicted without
a complete enumeration of structures. Essentially, this is possible
because we have found that the designability of a structure is closely
connected to its pattern of surface exposure.  Structures with large
variation in surface exposure are likely to be highly designable,
structures with more uniform surface exposure are not.  The higher the
designability of a structure, the more clearly differentiated are its
surface and core.  Because the variation in surface exposure of a
structure is independent of all other structures, designability can be
estimated structure by structure without the need for complete
enumeration.

One implication of this result is that candidates for highly
designable {\it long-chain} structures can be identified simply from
their surface-exposure patterns. This approach avoids the need for a
complete enumeration of structures. It is therefore computationally
feasible to consider much longer peptide chains, with a greater
variety of backbone conformations.  We demonstrate the efficiency of
this approach by generating backbone configurations of up to $N=40$
monomers.  For these lengths, complete enumeration of structures would
be impractical. Instead, we generate a relatively sparse sample of
structures. From among the sample it is straightforward to select the
candidate highly designable conformations as those with the most
clearly delineated surfaces and cores. An effectively equivalent
procedure is to calculate designability allowing only the structures
in the sparse sample to compete for sequences. In either case,
successful prediction of designability relies on its close relation to
surface-exposure variation -- a property of individual
structures. There is one important caveat to this point -- two
structures with very similar patterns of surface exposure will compete
for structures often in a ``winner-take-all'' fashion.  The
implications of this for design are discussed in Sec. III on random
sampling.

The results of the off-lattice model motivate us to consider the
surface-exposure pattern of natural protein structures. We report a
study of the surface-exposure pattern of backbones of up to length
$N=75$ from the Protein Data Bank. For small proteins, which are often
stabilized by disulfide bonds and salt bridges, there is often no
clearly delineated core.  In contrast, for large proteins the core is
uniformly well defined with little variation from structure to
structure.  The most highly designable configurations generated using
our sampling technique have patterns of surface exposure that fall
within the range of naturally occurring proteins.

\section{Models and Methods}

The designability of a structure is a measure of how many sequences
``fold'' into that structure in relation to all other competing
structures. To precisely determine designability requires generating a
comprehensive set of structures, which then compete as possible
lowest-energy states for amino-acid sequences.  It is only truly
feasible to generate a complete set of model structures for short
polypeptide chains. For larger chains, say with N $>$ 30 monomers, it
is not currently possible to enumerate structures.  However, it will
be shown below that it is possible to {\it estimate} the designability
of structures without complete enumeration.  To find the best means of
estimating designability, we study short chains ($N=15,17,\ {\rm and}\
19$) for which designability can be precisely determined within an
off-lattice model.  This section reviews our model for obtaining the
designabilities of short-chain polypeptide structures.  In the next
section, we show that the designability of a structure within this
model can be estimated from its surface-exposure pattern.

{\it Off-lattice model} -- Our method of generating structures is
closely related to the discrete-angle models introduced by Park and
Levitt.\cite{Park95,Miller00} For short polypeptide chains of
$N=15,17, {\rm and}\ 19$ monomers, a ``complete" set of backbones is
generated using a fixed set of three dihedral $(\phi,\psi)$
angles.\cite{Park95} For this particular study, we employ one
angle-pair $(-60,-50)$ from the alpha-helical region of a Ramachandran
plot, and two angle-pairs $(-140,150)$ and $(-65,125)$ from the
beta-strand region. The complete set of $3^N$ backbones is generated
with these angles.

To restrict our consideration to self-avoiding structures, we
introduce ``side groups'' by hard spheres of radius $r_{\beta} =
1.9\AA$ centered on the C$_\beta$ positions.  Self-avoidance is taken
into account by discarding all structures with overlapping spheres.
The percentage of self-avoiding structures out of the possible $3^N$
structures was found to be 42\% for $N=15$, 36\% for $N=17$ and 31\% for $N
= 19$.

An example of a structure generated using the three angle-pairs
and with self-avoiding spheres centered on the C$_\beta$ positions 
is shown in Fig.~\ref{fig1p}.

{\it Hydrophobicity model} -- There is considerable evidence that
hydrophobic forces are primarily responsible for the folding of an
amino-acid sequence into a particular structure.\cite{Kauzmann59,Dill90,Li97}
The hydrophobicity of each type of amino acid can be determined
experimentally.\cite{Nozaki71,Levitt76,Roseman}
Those which are more hydrophobic are energetically favored to reside
in the core of the folded protein, where there is low exposure to
water. In a given folded protein, a hydrophobic energy can be
assigned to each particular amino acid according to its hydrophobicity
and its exposure to water.

To determine the hydrophobic energy of an amino-acid sequence folded
into one of our model structures it is necessary to determine the
exposure of each residue along the backbone. As described above, hard
spheres are placed on each $C_\beta$ position, and the surface
exposure of these spheres to water is evaluated. This is done using
the method of Shrake and Rupley\cite{surface} which determines how
much of a sphere centered on a $C_\beta$ position is exposed to a
water molecule, which is represented as a sphere with a radius of
$1.4$ Angstroms.  We use the notation that the $j^{th}$ residue of the
$\alpha^{th}$ structure has accessible surface area $a^\alpha_j$. The
sum of these surface areas gives the total residue accessible surface
area for a given structure. As a screen, we use this quantity to
remove those structures which have too much surface exposure and thus
are too open to be stable folds. In practice, we reduce the
representative set to approximately $5000$ structures with the least
exposed surface area.\cite{discarded}

These remaining structures are ``compact'' in that even for the small
peptide chains used in this study ($N=15,17,\ {\rm and}\ 19$) there is the
beginning of the formation of a core which is inaccessible to solvent. We
then normalize the surface area of the remaining compact structures using
the following normalization condition, 
\begin{equation}
\tilde{a}^{\alpha}_j = \frac{a^{\alpha}_j} {\sum_j a^\alpha_j}.  
\label{eq:norm} 
\end{equation} 
The motivation behind this normalization procedure is that we wish the
remaining compact structures to all be {\it equally} compact -- the normalized
structures all have the same total surface exposure ($\sum_j
\tilde{a}^\alpha_j = 1$). This is in line with lattice studies where
all structures are equally compact.\cite{Govin95,Li96,Li98}
The normalization eliminates the need for an overall compactification 
energy in the energy function used below since all structures are
equally compact. 
Physically, the use of equally compact structures
accounts for the tendency of each structure to relax to its best
packed equivalent. 

With the residue-by-residue surface areas of a set of compact,
self-avoiding structures in hand, all that is needed is a
hydrophobic-energy function to associate these structures with
amino-acid sequences.  We find it convenient to assign a {\it
polarity} between $0.0$ and $1.0$ to each amino acid, with $0.0$ being
highly hydrophobic and $1.0$ being highly polar. Our notation is that
the $j^{th}$ amino acid of a sequence $\beta$ has polarity
$p^\beta_j$. In our model, the energy of this amino-acid sequence when
folded into the $\alpha^{th}$ structure is\cite{Li98,Eisenberg86,Buchler00}
\begin{equation}
E^{\beta,\alpha} = -\sum_j p^\beta_j \tilde{a}^\alpha_j . \label{eq:E}
\end{equation}
For a given sequence, the lowest-energy structure is the one that
minimizes this energy. Note that since all structures have the
same total exposed surface ($\sum_j \tilde{a}^\alpha_j = 1$),
a sequence will have lowest energy on the structure that best
matches its {\it pattern} of hydrophobicity  -- more hydrophobic at core
sites, more polar at surface sites -- independent of the absolute
hydrophobicity or polarity of the sequence.

{\it Designability} -- The designability of a given structure is
defined as the number of sequences with that structure as a unique
lowest-energy conformation.\cite{Li96} We assess the designabilities
of structures by evaluating the energy (\ref{eq:E}) of a large number
of random sequences of polarities on all the structures in the
representative set.  Each sequence of polarities $p_j$ is generated as
a string of $N$ random real numbers between $0.0$ and
$1.0$. Consistent with a previous study,\cite{Miller00} we report in
the next section that most structures are the lowest-energy
conformations of only a few or no sequences, and hence these
structures have low designability. Only a small fraction of structures
are highly designable. 

\section{Results and Discussion}

We now examine the factors which influence a structure's
designability. What causes a structure to be the lowest-energy state
of many sequences within our hydrophobic model?  We show below
that the {\it variance} of a structure's surface-area pattern is an
important quantity in determining designability.

{\it Predictors of designability} --
In our model, the energy of a sequence folded into a particular
structure is given by Eq.~(\ref{eq:E}). Therefore, the only property
of a structure which influences energy is the structure's vector of
solvent-exposed surface areas $\vec{\tilde{a}}= \tilde{a}_1,\ldots,
\tilde{a}_N$.  Moreover, because of the normalization condition,
Eq.~(\ref{eq:norm}), all such vectors reside on an $N$-dimensional
hyper-plane ({\it e.g.} for a chain of length N = 3, the vectors would
reside on the plane $\tilde{a}_1 + \tilde{a}_2 + \tilde{a}_3 = 1$).
The vectors for all structures can be decomposed into a constant
component $\vec{n}=(1/N, 1/N,\ldots,1/N)$ normal to this hyper-plane plus a
variable in-plane component (see Fig.~\ref{fig1}).  We denote a
structure's in-plane component by $ \vec{r} = \vec{\tilde{a}} -
\vec{n}$.  For a given sequence, the relative energies of structures
depend only on these $ \vec{r}$'s, as can be seen by rewriting
the hydrophobic energy as
\begin{equation}
E^{\beta,\alpha} = - \sum_j p^\beta_j  r^\alpha_j 
- \frac{1}{N}\sum_j p^\beta_j, \label{eq:Etwo}
\end{equation}
where the last term is structure independent.

The first term in the energy (\ref{eq:Etwo}) is the negative of an
N-dimensional dot product between the polarity vector $\vec{p}$ of the
sequence and the in-plane component $ \vec{r}$ of the structure.  The
lowest-energy structure is the one for which this dot product is the
greatest. The vector $\vec{p}$ can also be written in terms of a
component parallel to the normal vector $\vec{n}$ of the hyper-plane
and a component that lies in the hyper-plane.  For a given sequence,
the lowest-energy structure will be the structure that has the
greatest projection of its in-plane component $ \vec{r}$ onto the
in-plane component of the vector $\vec{p}$. Hence, structures that lie
the farthest out from the ``origin of the hyperplane'', $\vec{n}$, on
this plane will tend to be lowest-energy structures for the most
sequences ({\it e.g.}, structure ``S'' shown in
Fig.~\ref{fig1}). Distance from the origin $\vec{n}$ on the
hyper-plane is therefore expected to be an easy-to-calculate predictor
of designability.  For a given structure this distance is
\begin{equation}
 r = \sqrt{\sum_j ( r_j)^2}. \label{eq:var}
\end{equation}
Note that $r^2/N$ is the variance of a structure's
residue-by-residue exposed surface area.  From a physical point of
view, the in-plane distance $ r$ is a measure of how much
variation there is in a structure's exposed surface area compared to
uniform exposure to solvent. Structures that have large values of
$ r$ have well differentiated core and surface sites. 

However, designability is determined by more than just $ r$.  By
definition, to be highly designable, a structure must be the
lowest-energy state for a large number of sequences $\vec{p}$. For
each of the structures with large $ r$ there exists a kind of
``hyper-cone'' of sequences $\vec{p}$ for which it is the
lowest-energy state\cite{Li98,Buchler00} (this is shown schematically
by the shaded area in Fig.~\ref{fig1}).  The volume of this cone, and
hence the designability of the structure, depends on the density of
competing structures around it. This suggests that structures that lie
farthest from other structures on the hyper-plane will be most
designable. For example, a structure that is not the farthest out in
its own direction will tend to be less designable because a farther
out structure will be lower in energy for all sequences lying in the
same direction.  Hence, an improved predictor for designability is the
distance of a structure from the center of the distribution of
structures.\cite{Buchler00} We denote this distance from the mean by
\begin{equation}
\sigma = \sqrt{ \sum_j \sigma_j^2},
\label{sigmaone}
\end{equation}
where
\begin{equation}
\vec{\sigma} =   \vec{r} - \langle \vec{r}\,\rangle,
\label{sigma}
\end{equation}
and where $\langle \vec{r}\,\rangle$ is the mean of the distribution of
exposure vectors in the plane,
\begin{equation}
\langle \vec{r}\,\rangle  = \frac{1}{S_{\rm tot}}
\sum_{\alpha = 1}^{S_{\rm tot}}
 \vec{r}^{\, \alpha}, \label{meanr} \end{equation} with $S_{\rm tot}$ the
total number of structures in the set. One can determine if a structure is
the farthest out in its own direction by simply projecting all of the
other $\vec{\sigma}$'s onto its own $\vec{\sigma}$.  Structures that have
a large distance from the mean and also lie the farthest out in their own
direction are shown below to be highly designable. It has been previously
shown in lattice models that the designability of a structure is 
inversely correlated with 
the density of other structures in its local
neighborhood.\cite{Li96} However, to generate enough structures to
adequately sample local densities essentially requires complete
enumeration of structures. In contrast, the quantities $\vec r$
and $\vec \sigma$ depend only upon a structure's global position
within the space and require sampling of relatively few structures to 
compute.  Fig.~\ref{fig1} illustrates
the quantities of interest,$ \vec{r}$ and $\vec{\sigma}$, for a particular
structure that lies far from the origin on the hyper-plane. We now examine
how these quantities correlate with designability for some specific cases.

{\it Enumeration studies of 15, 17, and 19mers} --
The complete set of all self-avoiding compact structures was generated
using the three angle set described in Sec. II for lengths 
$N = 15,17,\  {\rm and} \ 19$. 
For each set of structures at a given length $N$, we
evaluated designability using the enumeration method described in Sec.~II, and
ranked the structures from highest to lowest designability. In
Fig.~\ref{fig2}, the histogram of designability for the 17mer is
shown. Consistent with other studies of designability, the histogram
has an exponentially decreasing tail of highly designable
structures.\cite{Li96,Miller00} Most of the structures in the
representative set have low designability, whereas only a few are
highly designable.

The distances $ r$ and $\sigma$ were also computed for all
structures.  The difference between the in-plane distance $ r$
and the distance from the mean $\sigma$ arises from the fact that the
mean exposure vector in the plane $\langle  \vec{r}\,\rangle$ is not
zero.  In Fig.~\ref{fig3a}, the mean vector $\langle
\vec{r}\,\rangle$ for the $N=17$ case is shown. From the plot it is clear that
the ends of a 17mer are on average more exposed to solvent than the
central portion.  For the other lengths studied the same behavior was
found, namely, averaged over the representative set, 
the ends of the structures tend to be more exposed
than the central portion.

Before looking at how designability correlates with $\sigma$ and
$ r$, we briefly show how they relate to a structure's
compactness. In Fig.~\ref{fig3b}, a histogram of the in-plane
distances $ r$ is shown for {\it all} self-avoiding 17mers
(solid line) and the 5000 most compact 17mers (dashed line). The
figure shows that our screen of structures for compactness removes
most structures which have low values of $ r$, but does not
remove those that have high values of $ r$. Hence structures
that have a large in-plane distance $ r$, {\it i.e.} large
surface-exposure variation, are also compact.

Fig.~\ref{fig4} shows the correlation between the in-plane distance
$ r$ and the distance from the mean $\sigma$, and how these
relate to designability for the $N=17$ case. There is a clear
correlation between the two distances $\sigma$ and $ r$, but
$\sigma$ is a better predictor of designability. The top 50 designable
structures are shown as black circles in Fig.~\ref{fig4}. Most of the
top 50 designable structures have values of $\sigma > 0.075$, and only
a few less designable structures have $\sigma$ values this high.  In
contrast, only about 10 out of the 50 topmost 
designable structures have values of $ r
> 0.075$, and for values less than this there is a mixture of
designable structures with less designable ones. Hence, high $\sigma$
is a better discriminator of high designability than high $ r$.
This difference between $\sigma$ and $ r$ reflects the fact that
distance from the mean $\sigma$ better identifies the structures which
are outliers from the distribution, and are hence likely to have high
designability.
We have thus found a quantity $\sigma$, determined purely from
a single structure's surface-exposure pattern, that can be used
to identify highly designable structures.

The implication of having a quantity that can estimate designability
from the properties of a single structure is that enumeration of a
large set of competing structures is not necessary.  This lifts the
severe computational constraint that enumeration places on the size
and complexity of structures that can be considered.  In the remainder
of this section, we study in more detail the relation between
surface-exposure distance from the mean $\sigma$ and the designability
of structures. Our attention is focused on the possibility of
identifying highly designable structures within a random sample, using
either designability within the sample or distance from the mean
$\sigma$.

Figure~\ref{fig5} shows the designability versus $\sigma$ for chains
of length $N=15, 17,\ {\rm and}\ 19$.  Over $10^6$ sequences were
generated to determine the designability of structures in each case.
The structures for each chain length were binned according to $\sigma$
and the average designability of the structures in each bin is
plotted. The correlation between $\sigma$ and designability is
clear, but there is an important caveat. We have only plotted the
designability of structures which are the farthest from the mean in
their own direction -- this means that there are no other structures
whose $\vec{\sigma}^{\, \alpha}$ has a greater projection onto the
given structure's direction $\vec{\sigma}$, {\it i.e.} $ \vec{\sigma}
\cdot \vec{\sigma}^{\,\alpha} < |\vec{\sigma}|^2$. for all other
structures $\alpha$.  In this way, we have plotted only the 
``winners'' 
in the winner-take-all competition for sequences that occurs when
two or more structures have very similar patterns of surface
exposure. In part, this procedure is justified to select only one
member from every family of geometrically closely related
structures.\cite{Miller00} Fig.~\ref{fig5b} illustrates the effect of
this winner take all competition for the $N=17$
case. Fig.~\ref{fig5b}(a) is a plot of designability versus $\sigma$ for
all structures, while Fig.~\ref{fig5b}(b) is 
a plot of designability versus $\sigma$ for the ``winning" structures,
which are the farthest from the mean in their own direction. 
It can be seen that there are a
large number of structures which have low designability despite having
a large distance from the mean. Their designability has been reduced
due to competition with a structure which is farther out on the
hyper-plane. However, the structures with the largest values of 
distance from the mean $\sigma$ are all highly designable.

In Fig.~\ref{fig5}, the marked vertical lines on each graph indicate
the values of $\sigma$ for the 10th, 100th, and 1000th ranked
structures according to $\sigma$ in the {\it entire} set of compact
structures. To highlight the significance of this in regards to random
sampling, consider the following: if only $0.1$\% of the 17mer
structures were sampled, it would still be expected that the 1000th
ranked structure would occur in the sample. From the graph of the
17mer, even the 1000th ranked structure still has a reasonably high
designability in that, on average, it is the ground state for a few
hundred sequences. It is interesting to note that the designability
versus $\sigma$ curves become steeper as chain length $N$
increases. Hence, it is reasonable to conclude that for larger protein
structures, $\sigma$ {\it improves} as a predictor of designability.

To emphasize how a structure's location within fold space influences
its designability, in Fig.~\ref{fig6} we present histograms of the
number of nearby structures for the 1st, 100th, and least 
designable ``winner'' structures.  To make these plots, we calculated
the projections of all $\vec{\sigma}$ vectors onto the selected
structure's $\vec{\sigma}$ vector. The projections were normalized by
the magnitude of the selected structure's $\vec{\sigma}$. The
histograms show the number of structures that have a given projection
onto the chosen structure.  Fig.~\ref{fig6}(a) shows the histogram for
the most designable 17mer structure. There is a large distance between
it and the next nearest structure. However, for the less designable
structures in Figs.~\ref{fig6}(b) and (c) this distance is much
shorter. 
In fact for the structure in Fig.~\ref{fig6}(c), there are
two other structures whose projections lie so close to the chosen
structure as to fall into its own bin, of size 0.1. 
These near neighbors in fold space
compete for sequences and are responsible for the low designability value.

{\it Random sampling for 17mers} -- To show that highly designable
structures can be identified within a sparse random sample we
performed the following test.  We randomly selected a set of 500
structures out of the complete set of 5000 compact 17mers. A
designability calculation was then done for this small random sample.
In Fig.~\ref{fig7}(a), we plot the designability calculated using the
complete set versus the designability in the random sample, for those
structures that were the farthest out in their own direction. The
correlation is good, with the highest designability structure
correctly identified. The essential reason underlying the good
agreement is the close correlation between designability and $\sigma$
(cf. Fig.~\ref{fig5}) combined with the fact that $\sigma$ for the
random sample is effectively the same as $\sigma$ for the complete
set. This last relation is shown in Fig.~\ref{fig7}(b), where we have
re-evaluated the $\sigma$'s using the new mean $\langle\vec{r}\,\rangle$
of the random sample.  The logic of random sampling is simple -- we
can identify highly designable structures from their large $\sigma$
values, and only a small sample is required to calculate these
$\sigma$'s.  In fact, the designability for the complete set can be
directly estimated from the $\sigma$ values in the random sample, as
shown in Fig.~\ref{fig7}(c). The correlation is slightly better using
the designability calculated within the random sample, as shown in
Fig.~\ref{fig7}(a), but the practical consideration of avoiding
designability calculations may in some cases favor the direct use of
$\sigma$ values.

{\it Random Sampling of Long-Chain Structures }--
We now show how to find candidates for highly designable
long-chain structures by random sampling.  For backbone
configurations of length $N > 30$, complete enumeration of structures
is infeasible.  However, according to the results of the previous
section, one can randomly sample long-chain structures, evaluate their
surface exposures, and use the variation $\sigma$ to estimate which
ones are likely to be highly designable.  Without the constraint of
enumeration, one is free to consider more complex backbone
configurations, for example, using a larger number of $(\phi,\psi)$
angle-pairs. In addition, the sampling of structures can be biased to
favor configurations with realistic secondary structural elements.

To generate long-chain structures, we employed a set of four ($\phi,
\psi$) angle-pairs.  The four pairs of angles were taken from high
density regions of a Ramachandran plot: two pairs from the
alpha-helical region, one from the beta-strand region, and the last
from the left-handed alpha-helical region.\cite{angles2} Structures
were generated by randomly selecting a pair of angles, weighted
equally, and then randomly selecting proceeding angles from a
transition table (Table 1), so that alpha-helical angles tend to
follow alpha-helical angles, and beta-strand angles follow beta-strand
angles. The matrix of transition probabilities in Table 1 was adapted
from an analysis of transition probabilities between $(\phi,\psi)$
pairs of naturally occurring protein structures.\cite{millerangles}
The transition probabilities involving the left-handed alpha-helical
angle were altered to include more turns to generate compact 30mers
and 40mers for this study. As before, hard spheres were centered on
each $C_\beta$ position and self-avoidance was enforced by eliminating
structures with overlapping spheres. We found that the use of the
transition probabilities dramatically reduced the generation of
self-intersecting structures.

We randomly sampled both 30mer and 40mer structures using the above
procedure. For each length, the 500 most compact, self-avoiding
structures generated in approximately two days of computing time on a
600 MHz PC were retained. Both the in-plane distance $ r$ and
the distance from the mean $\sigma$ were evaluated for the 500
structures in each random sample. Fig.~\ref{fig8} shows the top two
30mers and top two 40mers ranked by $\sigma$. According to the results
of the previous section, these structures are our best candidates to
be highly designable.  We compare their values of $\sigma$ to those of
naturally occurring structures in the section below, and show that
their structural characteristics are consistent with naturally
occurring proteins.

An important caveat to the random sampling approach is that there
could exist unsampled structures, with very similar patterns of
surface exposure, that would compete for sequences with our top
structures. Competing chain configurations that are {\it
geometrically} similar can be considered as fluctuations of a single
structure.\cite{Miller00} However, the possibility of geometrically
dissimilar structures with similar surface-exposure patterns, is an
unavoidable uncertainty associated with random sampling. This
competetition for sequences between geometrically dissimilar 
structures has been recently studied in lattice models.\cite{Kaya}

{\it Surface-Exposure Patterns of Naturally Occurring Structures }--
We now examine the surface-exposure patterns of naturally occurring
proteins. From the Protein Data Bank (PDB), we selected  groups of
unrelated structures of fixed length $N$, with $N=25\ldots 75$, and
extracted their backbones.  We then positioned uniformly sized spheres
on the $C_\beta$ positions, and evaluated surface exposure exactly as
was done for the small chains studied above.  Both the in-plane
distance $ r$ and the distance from the mean $\sigma$ were
evaluated for the 71 natural occurring structures in the set.

Fig.~\ref{fig9}(a) shows the length dependence of $r$ for the selected
set of natural protein structures using spheres with radius $1.9$
Angstroms on the C$_\beta$ positions. (We have chosen to plot $r$
rather than $\sigma$ since $\sigma$ depends on the evaluation of
$\langle \vec{r} \,\rangle$ which has a large error due to the small size of
the sample). For small chains there is a broader variation than for
the longer chains. This can be attributed to the fact that small
proteins are often stabilized by disulfide bridges, rather than by the
formation of a hydrophobic core.\cite{Creighton93} In particular, for
the $N=25$ proteins, most structures had ill-defined cores, hence the
lower values for the variance $r$. For larger proteins the
distribution is narrower. This suggests that for larger proteins the
hydrophobic force plays a more consistent role in creating a
well-defined hydrophobic core.
The average surface-exposure variation $r$ decreases slightly 
with chain length. This could be anticipated from our normalization
procedure: The total surface exposure for compact structures 
grows as $N^{2/3}$. If the variance of surface exposure of 
individual $C_\beta$ spheres stays fixed, then the normalization
implies $r_i^2 \sim N^{-4/3}$, and so $r \sim N^{-1/6}$
according to Eq.~\ref{eq:var}. 

The highest $r$ structures from the random sampling study above are shown
as filled circles in Fig.~\ref{fig9}(a). These structures lie at the high
end of the variances of the naturally occurring proteins. This result is
encouraging as it suggests that our best randomly generated structures
share similar properties to real protein folds. However, the naturally
occurring structures tend to be more open than the randomly generated
structures. Hence, using small uniform spheres on the C$_\beta$ positions
overestimates the accessible regions of the natural structures. In
Fig.~\ref{fig9}(b), the in-plane distance of the selected 40mers from the
PDB is shown as a function of side-chain sphere radius. Using larger
spheres increases the variance and hence in-plane distance of the natural
structures. Nevertheless, the top randomly sampled 40mer structure still
falls within the mid to high range of the variance even when more
realistic side-chain sphere sizes are used for the naturally occurring
structures.

\section{Conclusions}
In conclusion, we have shown that it is possible to estimate the
relative designabilities of protein structures based on their exposed
surface-area patterns, within an off-lattice model.  Specifically, the
{\it designability} of a structure -- defined as the number of
sequences with that structure as a unique lowest-energy state -- was
found to closely correlate with the surface-exposure variation of the
structure.  The ability to estimate designability from the properties
of a single structure makes it unnecessary to completely enumerate
structures. Instead, a sparse sample of structures can be generated,
and relative designability assessed from designability within the
sample, or directly from the surface-exposure variation of each
structure.  Random sampling, in turn, allows consideration of longer
chains with greater structural complexity. We have demonstrated the
random sampling approach to designability for 30mers and 40mers. Our
best candidates for highly designable structures were found to have
surface-exposure variations similar to those of naturally occurring
structures of the same size. Random sampling thus offers a promising
way to find highly designable long-chain structures for ab initio
protein design and also may be useful in the generation of decoys.


\begin{table}[!t]
\begin{center}{\bf Transition Probabilities between Dihedral 
Angle-Pairs}\\ 
\end{center}
\begin{tabular}{c|c|c|c|c}
$angle_1$$\backslash$$angle_2$  & $\alpha_1$ & $\alpha_2$ & $\beta$ & $L$ \\ \hline
$\alpha_1$ & 0.65 & 0.30 & 0.0 & 0.05 \\ \hline
$\alpha_2$ & 0.35 & 0.35 & 0.2 & 0.10 \\ \hline
$\beta$ & 0.1 & 0.1 & 0.8 & 0.0 \\ \hline
$L$ & 0.09 & 0.1 & 0.8 & 0.01 
\end{tabular}

\caption{Transition probabilities $\rm{Probability(column | row)}$ for 
successive angle-pairs for the four $(\phi,\psi)$ pairs used\cite{angles2}
to generate 30mer and 40mer structures}\label{tbl1}
\end{table}

\begin{figure}[!t]
\begin{center}
\caption{Depiction of a 19mer structure, constructed 
using the three $(\phi,\psi)$ angle-pairs described in the text.
The amino-acid side groups are modeled by self-avoiding spheres of radius
$1.9\AA$ centered on the C$_\beta$ positions.}
\label{fig1p}
\end{center}
\end{figure} 
\begin{figure}[!t]
\begin{center}
\caption{Schematic diagram of fold space. Black circles correspond to
vectors of exposed surface area $\tilde{ \vec{a}}$ for individual
structures.  Since the surface-area vectors are normalized, all lie on
a single hyper-plane. The vector for uniform exposed area $\vec{n} =
(1/N,\ldots,1/N)$ is the origin for all vectors on the
hyper-plane. The in-plane vector $ \vec{r} = \tilde {\vec{a}} -
\vec{n}$ is shown for one structure (labelled ``S'') with a large
$| \vec{r}|$, 
and thus a highly non-uniform exposed surface area. For
the same structure, the vector $\vec{\sigma} = \vec{r} - \langle
\vec{r}\,\rangle$, relative to the mean of the distribution
$\langle \vec{r}\,\rangle$, is also shown. Structures with 
surface-exposure patterns very different from the mean, and thus with large
values of $|\vec{\sigma}|$, are typically highly designable. 
The sequences which have ``S'' as their lowest-energy conformation,
and thus contribute to the designability of ``S'', are shown 
schematically by the shaded ``hyper-cone''.}
\label{fig1}
\end{center}
\end{figure}
\begin{figure}[!t]
\begin{center}
\caption{Histogram of designability for the 5000 most compact,
self-avoiding 17mers. The histogram has an exponentially decreasing
tail of highly designable structures.}
\label{fig2}
\end{center}
\end{figure}
\begin{figure}[!t]
\begin{center}
\caption{ Normalized exposed surface area versus
position of monomer on chain, averaged over the 5000 most compact,
self-avoiding 17mer structures. The dashed straight line corresponds to
the uniformly exposed structure $(1/17, 1/17,\ldots,1/17)$. Also shown is 
a typical normalized surface exposure pattern of a compact structure
(dot-dash)} \label{fig3a} \end{center}.
\end{figure}
\begin{figure}[!t]
\begin{center}
\caption{ Histogram of surface-exposure-variation magnitude $ r$
for all self-avoiding 17mers (solid line) and for the 5000 most
compact, self-avoiding 17mers (dotted line).}
\label{fig3b}
\end{center}
\end{figure}
\begin{figure}[!t]
\begin{center}
\caption{Plot of distance from the mean $\sigma$ against in-plane
distance $ r$ for the 5000 most compact, self-avoiding 17mers. 
Black circles correspond to
the 50 topmost designable structures and white squares to structures
that are less designable.  }
\label{fig4}
\end{center}
\end{figure}
\begin{figure}[!t]
\begin{center}
\caption{Plot of designability versus distance from the mean $\sigma$
for ``winner'' structures of length $N=15,17,\ {\rm and}\ 19$.  Error
bars indicate the uncertainty in designability for each bin.  The
vertical lines correspond to the 10th, 100th, and 1000th structures
ranked according to $\sigma$ in the entire sample.}
\label{fig5}
\end{center}
\end{figure}
\begin{figure}[!t]
\begin{center}
\caption{(a) Plot of designability versus distance from the mean
$\sigma$ for all structures of length $N=17$. (b) Plot of
designability versus distance from the mean $\sigma$ for ``winner''
structures of length $N=17$, {\it i.e.} those structures which lie the
farthest out in their own direction.}
\label{fig5b}
\end{center}
\end{figure}
\begin{figure}[!t]
\begin{center}
\caption{Histograms of the normalized projections of the
$\vec{\sigma}$'s of all 17mer structures onto the $\vec{\sigma}$ of the
(a) most designable, (b) 100th most designable, and (c) least designable
``winner'' structure. Note the change of $y$-axis scale in (c).}
\label{fig6}
\end{center}
\end{figure}
\begin{figure}[!t]
\begin{center}
\caption{(a) Plot of designability in the full set of 5000 compact,
self-avoiding 17mers versus designability calculated for a random
sample of 500 of these structures. Of the sampled structures, only
those that are the farthest out in their own direction on the
hyper-plane are shown. (b) Plot of surface-exposure distance from the
mean $\sigma$ in the full set versus $\sigma$ for the same random
sample of 500 structures.  (c) Plot of designability in the full set
versus $\sigma$ in the random sample.}
\label{fig7}
\end{center}
\end{figure}
\begin{figure}[!t]
\begin{center}
\caption{Top structures ranked according to surface-exposure 
distance from the mean $\sigma$ in sparse random
samples of 30mers and 40mers: (a),(b) top two ranked 30mers according
to $\sigma$; (c),(d) top two ranked 40mers according to $\sigma$. }
\label{fig8}
\end{center}
\end{figure}
\begin{figure}[!t]
\begin{center}
\caption{(a) Surface-exposure distance from the in-plane distance $r$ for
naturally occurring protein backbones in the Protein Data Bank for
$N=25-75$.  For each $N$, only structurally distinct proteins in the
Data Bank were selected. Also shown in black circles are the values of
$r$ for the top $30$mer and $40$mer from the randomly generated
structures shown in Fig.~\ref{fig8}. (b) The dependence of $r$ on
choice of 
side-chain sphere radius for the set of selected 40mers in the PDB. 
The value of $r$ of the top randomly generated 40mer 
is shown as a black circle.}
\label{fig9}
\end{center}
\end{figure}


\begin{references}
\bibitem{discarded} The discarded structures are too open
to have well defined cores and therefore have low 
designabilities, as discussed in Section III.
\bibitem{angles2} The four $(\phi,\psi)$ angles chosen were: $\alpha_1
= (-45,-55)$, $\alpha_2 = (-70,-20)$, $\beta = (-100,100)$, and
$L = (50,25)$ based on a clustering method given in
Ref.~\cite{millerangles}.
\end{references}
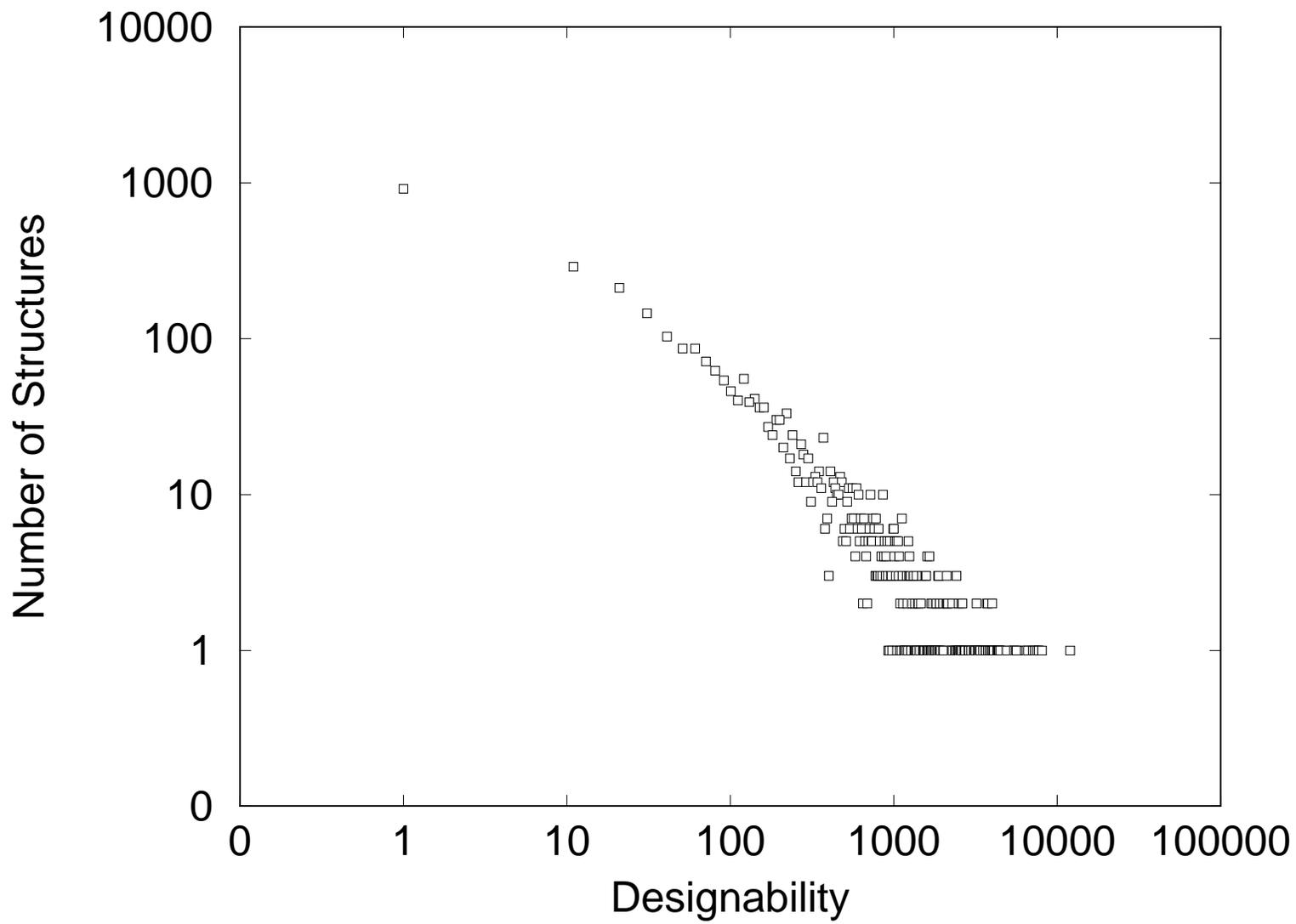
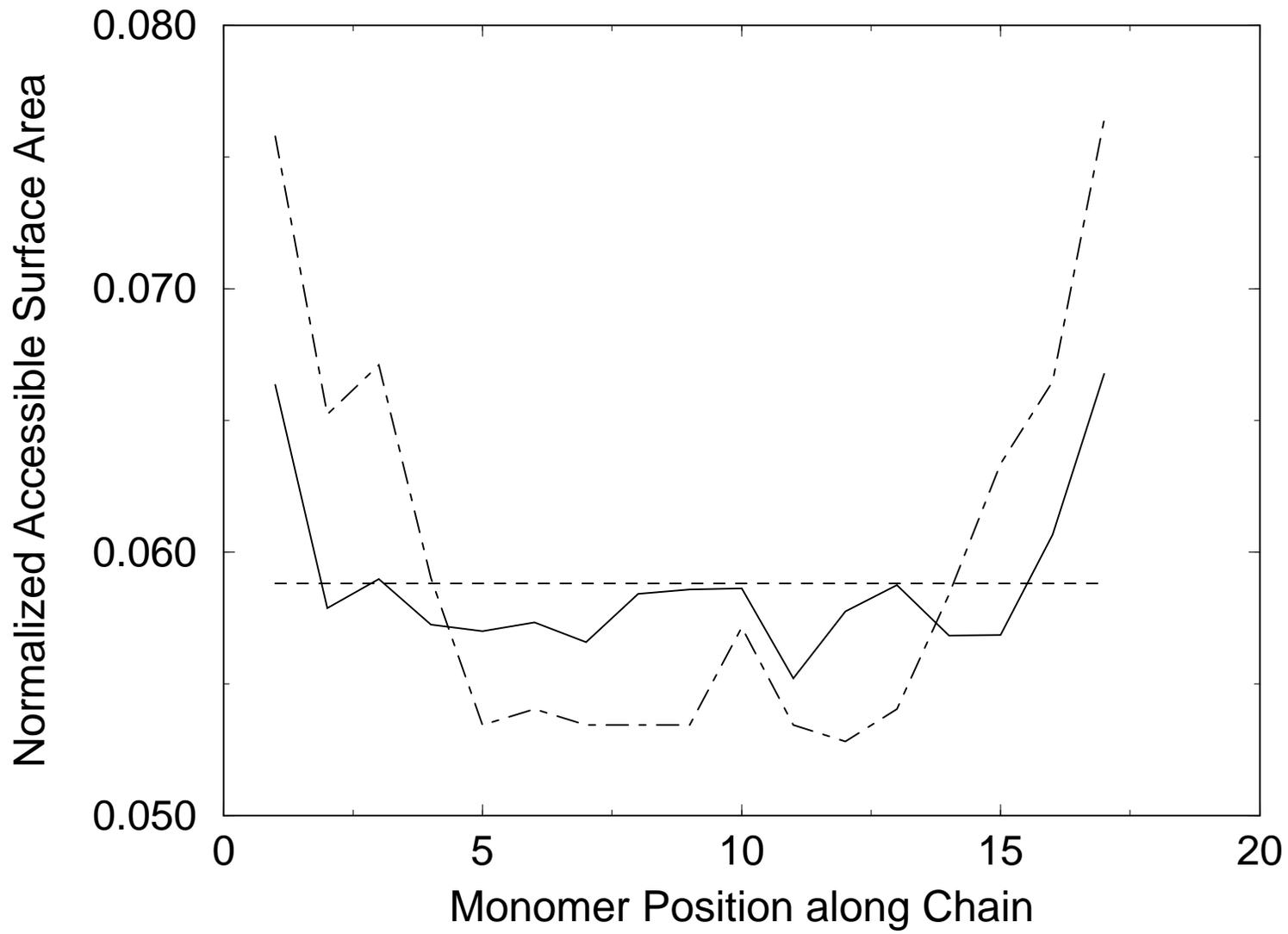
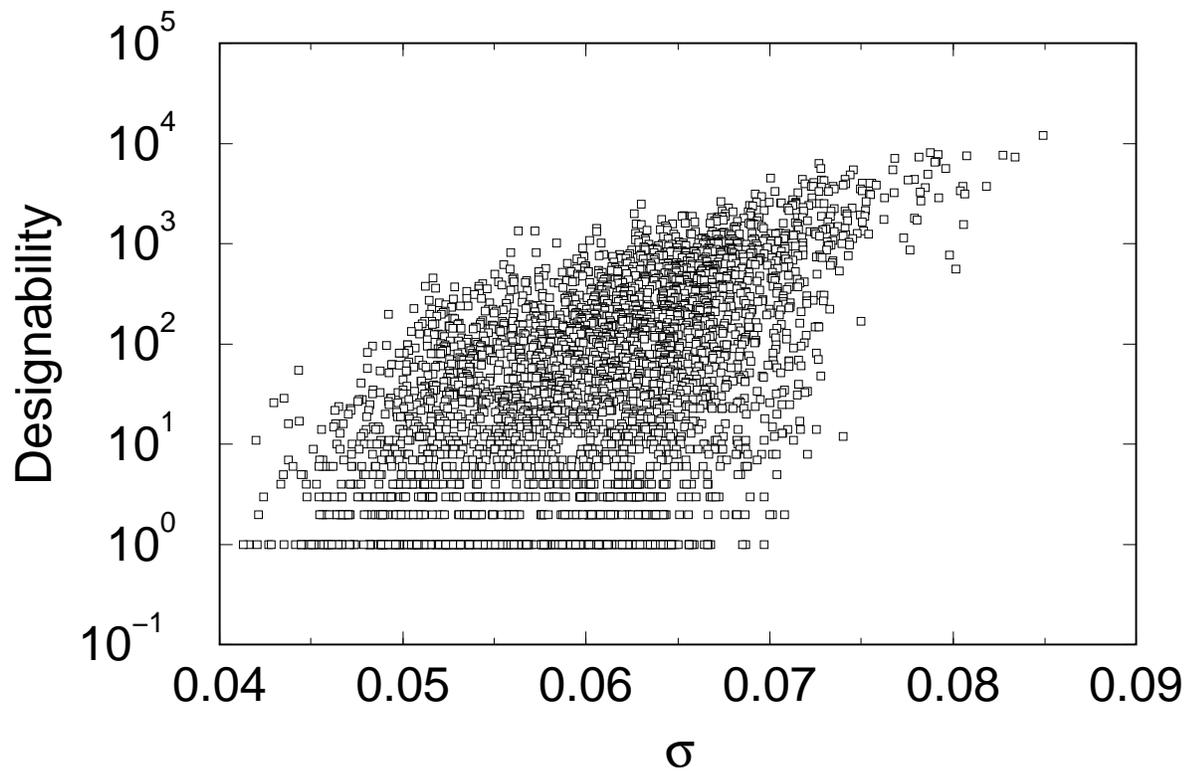
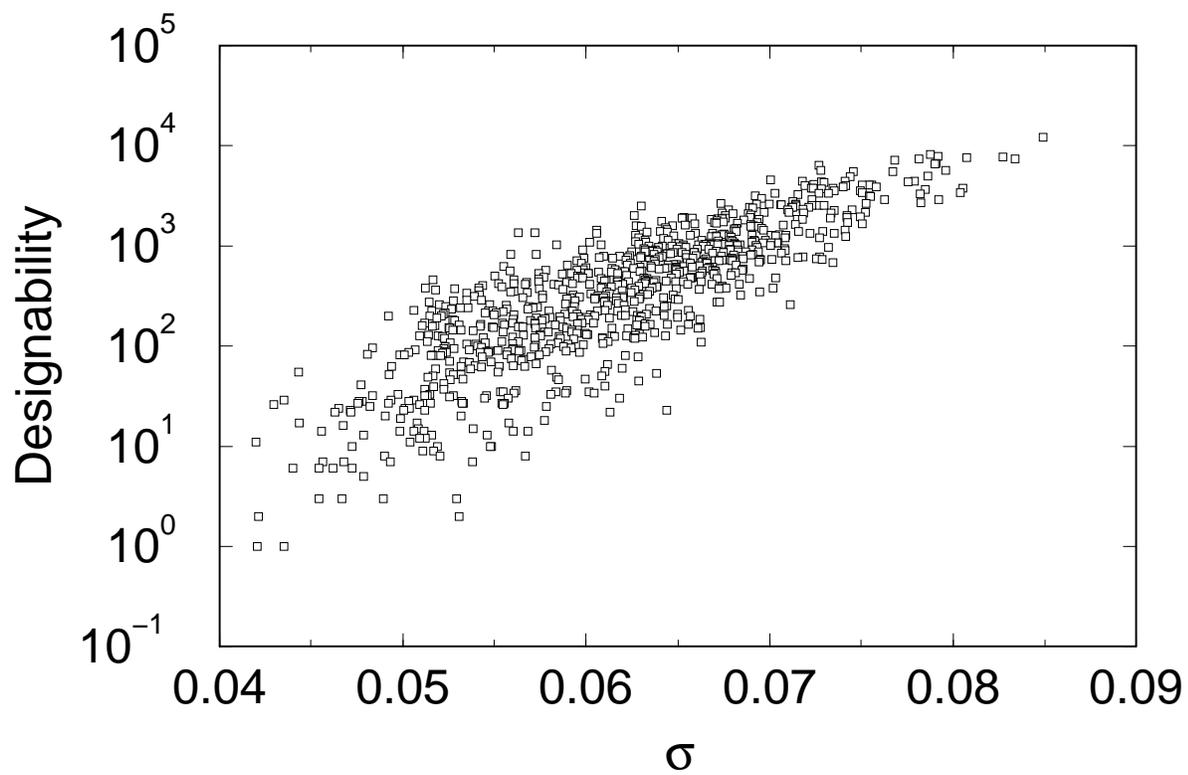
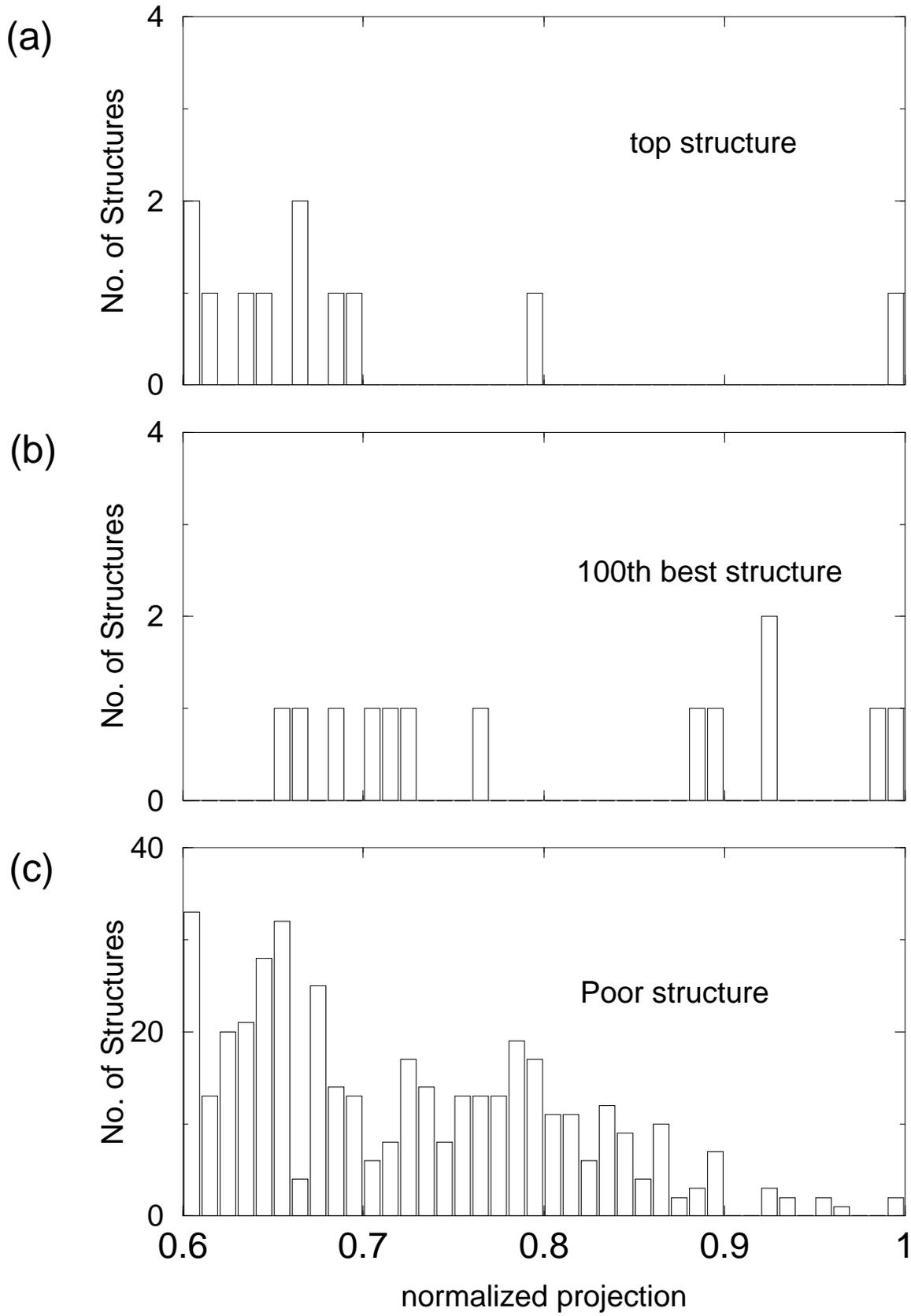
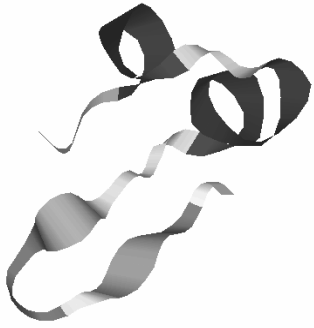
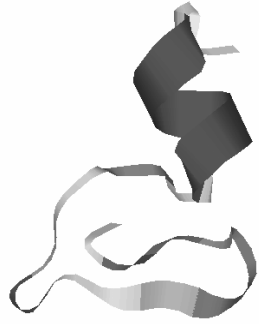
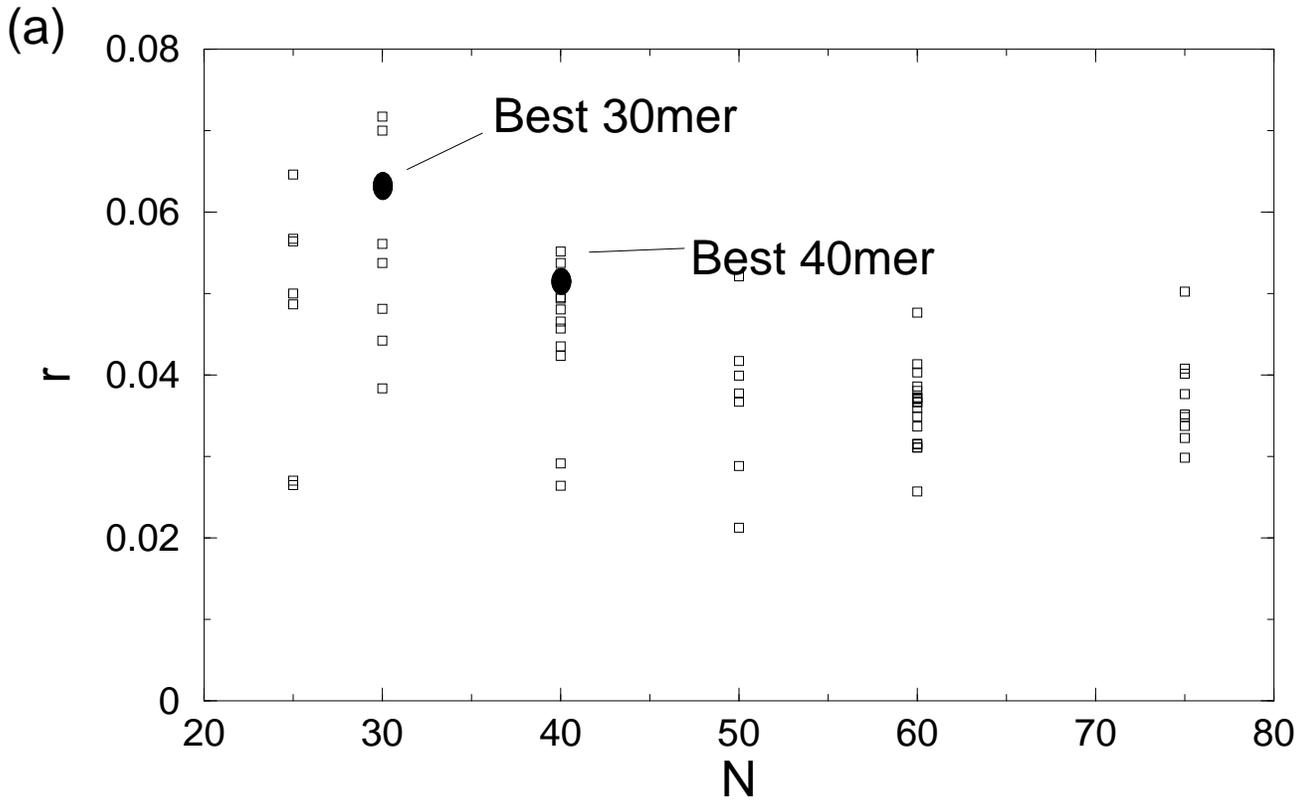
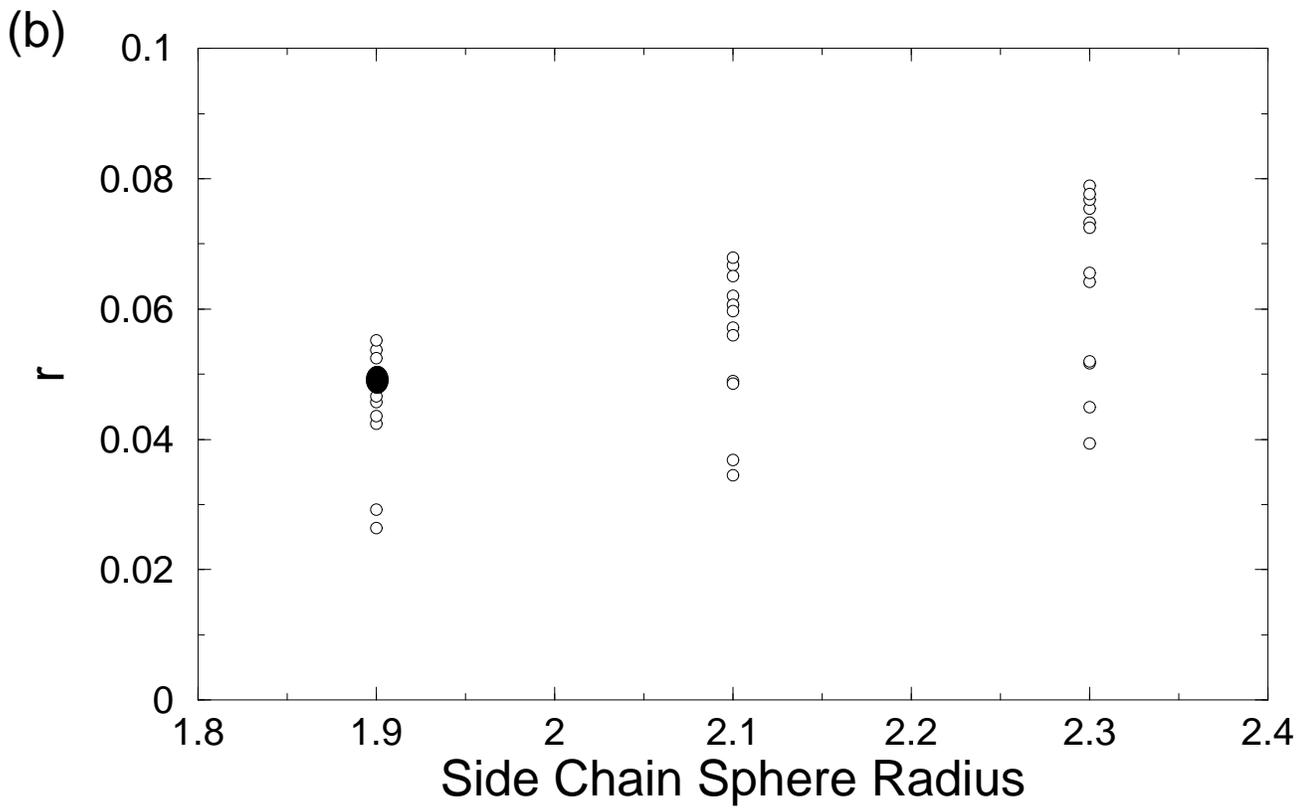
\end{document}